\documentclass[usenatbib,useAMS]{mnras}

 \usepackage{graphicx}
 \usepackage{amsmath}
 \usepackage{mathrsfs}
 \usepackage[varg]{txfonts}
 \usepackage{braket}
 \usepackage{cancel}
 \usepackage{color}

 \newcommand{\Gaia}{{\it Gaia}\ }
 \usepackage{soul}   

 \renewcommand\le\oldleq
 \renewcommand\ge\oldgeq

 \newcommand{\msun}{M_\odot}
 \newcommand{\Msun}{M_\odot}

 \renewcommand\pi\upi

  \title[Discovery of New Retrograde Substructures]{Discovery of New
    Retrograde Substructures: The Shards of $\omega$~Centauri?}

  \author[Myeong et al.]
          {G.~C.~Myeong$^1$\thanks{E-mail:~gm564,nwe,vasily,jls@ast.cam.ac.uk,
              koposov@cmu.edu}, N.~W.~Evans$^1$, V.~Belokurov$^1$,
            J.L. Sanders$^1$, S.E.~Koposov$^{1,2}$
            \\$^1$Institute
            of Astronomy, University of Cambridge, Madingley Road,
            Cambridge CB3~0HA \\$^2$ McWilliams Center for Cosmology,
            Department of Physics, Carnegie Mellon University, 5000
            Forbes Avenue, Pittsburgh, PA 15213, USA}

 \pagerange{\pageref{firstpage}--\pageref{lastpage}}\volume{000}\pubyear{2017}
 \date{version \today.}

\begin{document}
 \label{firstpage}
 \maketitle

\begin{abstract}
We use the SDSS-{\it Gaia} catalogue to search for substructure in the
stellar halo. The sample comprises 62\,133 halo stars with full phase
space coordinates and extends out to heliocentric distances of $\sim
10$ kpc.  As actions are conserved under slow changes of the
potential, they permit identification of groups of stars with a common
accretion history.  We devise a method to identify halo substructures
based on their clustering in action space, using metallicity as a
secondary check.  This is validated against smooth models and
numerical constructed stellar halos from the Aquarius simulations. We
identify 21 substructures in the SDSS-{\it Gaia} catalogue, including
7 high significance, high energy and retrograde ones.

We investigate whether the retrograde substructures may be material
stripped off the atypical globular cluster $\omega$~Centauri. Using a
simple model of the accretion of the progenitor of the
$\omega$~Centauri, we tentatively argue for the possible association
of up to 5 of our new substructures (labelled Rg1, Rg3, Rg4, Rg6 and
Rg7) with this event. This sets a minimum mass of $5 \times 10^8
\Msun$ for the progenitor, so as to bring $\omega$~Centauri to its
current location in action -- energy space. Our proposal can be tested
by high resolution spectroscopy of the candidates to look for the
unusual abundance patterns possessed by $\omega$~Centauri stars.
\end{abstract}
\begin{keywords}
{galaxies: kinematics and dynamics -- galaxies: structure}
\end{keywords}

\section{Introduction}

The spatial structure of the stellar halo has already been explored
using either multiband photometry from surveys like the Sloan Digital
Sky Survey and Pan-STARRS~\citep[e.g.,][]{Be06, Bell08, Sl14} or variable
stars such as RR Lyrae characteristic of old metal-poor stellar
populations~\citep{Wa09,Io18}. At least within heliocentric distances
of $\sim 30$ kpc and for declinations northward of $\delta
=-30^\circ$, the most prominent halo substructures in resolved star
density maps have now been identified by matched filter
searches~\citep{Ne15}.

Nowadays, we are so familiar with maps such as the ``Field of
Streams''~\citep{Be06} that we forget how surprising they really
are. Substructure identification in configuration space is grossly
inefficient compared to searches in phase space~\citep{Jo98,He99}.
Streams remain kinematically cold and identifiable as substructure in
phase space long after they have ceased to be recognisable in star
counts against the stellar background of the galaxy. Given what has
already been discovered with multiband photometry, the local
phase space structure of the stellar halo must be bristling with
abundant substructure.

 \begin{figure}
 \begin{center}
   \includegraphics[width=0.5\textwidth]{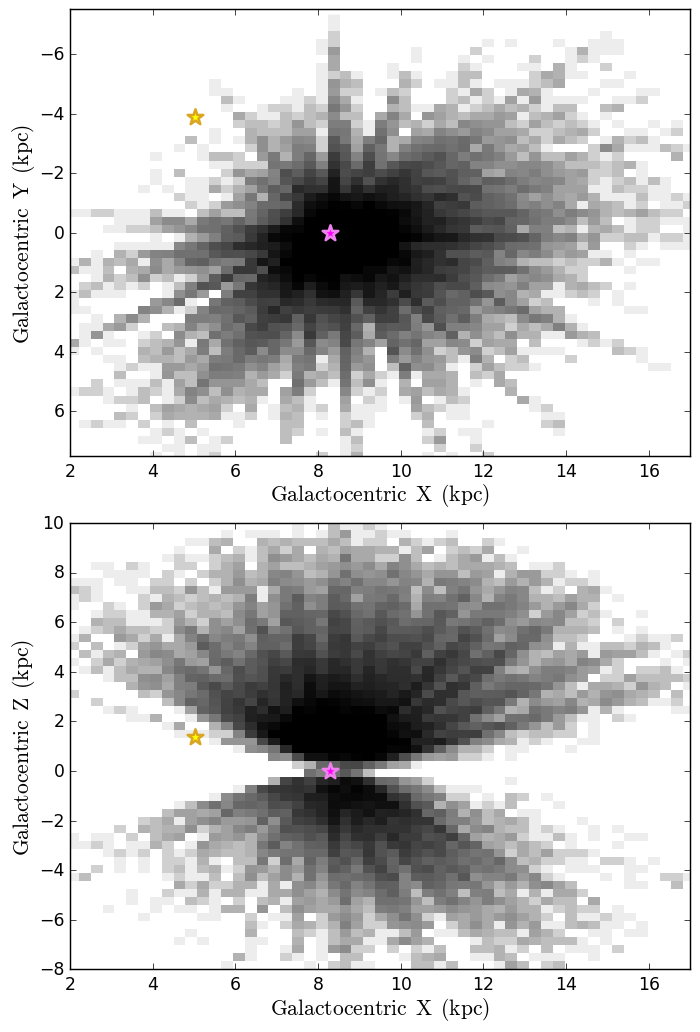}
 \end{center}
 \caption{Distribution of the stellar halo sample in the SDSS-{\it
     Gaia} catalogue in spatial coordinates projected onto the
   principal planes ($X,Y$) and ($X,Z$) in Galactocentric Cartesian
   coordinates ($X,Y,Z$). There are 62\,133 halo stars with full phase
   space coordinates and the sample extends out to heliocentric
   distances of $\sim 10$ kpc. The golden star in each panel
   represents the present position of $\omega$~Centauri, while the
   mauve star is the position of the Sun. Note that $\omega$~Centauri
   is at the low galactic latitude limit of the survey, so some of its
   debris may be missed.}
 \label{fig:figzero}
 \end{figure}

Astrometric satellites have the ability to transform this terrain.
Already using data from the {\it Hipparcos} satellite, \citet{He99}
identified 13 stars which form an outlier in the plane defined by two
components of angular momentum~\citep[see also][for later
  developments]{My18}. The first {\it Gaia} data release (DR1) in 2016
has already inspired two such searches. \citet{He17} used the
Tycho-\Gaia Astrometric Solution (TGAS) cross-matched with
RAVE~\citep{Ku17} to identify overdensities in ``integrals of the
motion space'', or energy and angular momentum space, which they
ascribed to halo substructure. \citet{My17} used TGAS cross-matched
with RAVE-on~\citep{Ca17} to search for halo substructure in action
space, identifying a subset of stars with large radial action. These
stars are all moving on highly eccentric orbits and are clustered in both
configuration space and metallicity, thus providing a
convincing candidate.

Crossmatches between TGAS and radial velocity surveys provide
catalogues of $\sim 2\,000$ halo stars largely within $\sim 1$ kpc of
the Sun. This is too parochial for studies of the stellar halo.  The
SDSS-\Gaia catalogue contains a much larger and deeper sample of $\sim
60\,000$ halo stars out to $\sim 10$ kpc. This catalogue was made by
recalibrating the Sloan Digital Sky Survey (SDSS) astrometric
solution, and then obtaining proper motions from positions in the
\Gaia DR1 Source catalogue and their recalibrated positions in
SDSS~\citep[see e.g.,][for more details] {De17,DeBoer18}. The
individual SDSS-\Gaia proper motions have statistical errors typically
$\sim 2$ mas yr$^{-1}$, or $\sim 9.48 D$ km s$^{-1}$ for a star with
heliocentric distance $D$ kpc. The SDSS-\Gaia catalogue is the natural
intermediary between \Gaia DR1 and the recently released \Gaia
DR2~\citep{GaDR2_sum}.

\citet{MyPre} recently provided new pictures of the Milky Way halo in
action space as a function of metallicity using a sample of
$\sim60\,000$ halo stars with full phase space coordinates present in
the SDSS-{\it Gaia} catalogue.  The comparatively metal-rich halo
($-1.9 <$ [Fe/H] $< -1.3$) is strongly retrograde at high
energies~\citep[see e.g., Figure 2 of ][]{MyPre}. By contrast, at
lower metallicities, there are very few halo stars that are retrograde
and high energy.  This is evidence of a considerable retrograde merger
or accretion event in the recent past~\citep[e.g.,][]{Qu86,No89}.

Here, we carry out a search for halo substructure in action space
using the SDSS-\Gaia catalogue. This is a modification of our earlier
search for halo substructure in velocity space~\citep{My18}. There are
a number of advantages to using actions.  Unlike integrals of motion,
actions preserve their invariance under slow
changes~\citep[e.g.,][]{Go80}.  They have often been suggested as the
natural coordinates for galactic dynamics~\citep[see e.g.,][]{Bi82},
in which of course the potential is evolving in time. \citet{He99}
first argued that fossil structures in the stellar halo may be
identifiable as clusters in action space. This idea has been tested
extensively with numerical simulations both in static analytical
potentials and in time-varying cosmological
potentials~\citep{He99,Kn05,Go10}.

 \begin{figure*}
 \begin{center}
\includegraphics[width=\textwidth]{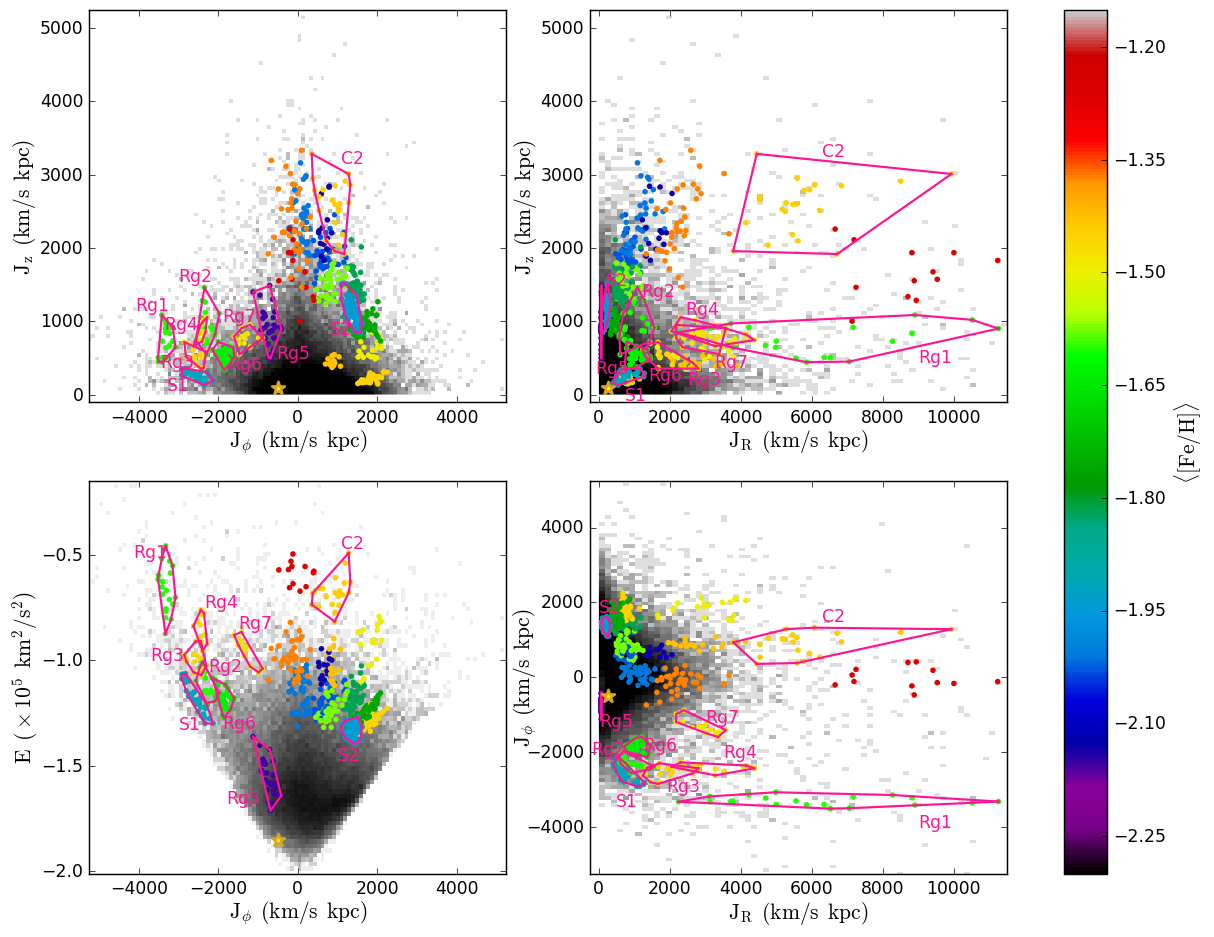}
 \end{center}
 \caption{Distribution of the stellar halo sample and substructure
   candidates in action -- energy space.  Top left:
   $(J_{\phi},J_z)$. Top right: $(J_z,J_R)$. Bottom left:
   $(J_{\phi},E)$. Bottom right: $(J_R,J_{\phi})$.  The 21 most
   significant substructures are colour-coded according to
   metallicity.  Previously found substructures (S1, S2, C2) and seven
   highlighted candidates (Rg1 -- Rg7) are further highlighted with a
   magenta outline. The golden star in each panel represents the
   present position of $\omega$~Centauri.}
 \label{fig:figone}
 \end{figure*}

The identification of substructures enables us to map out the
accretion history of the Milky Way. For example, \citet{My18} found two
prominent substructures in velocity space (S1 and S2) and used a
library of accreted remnants to estimate that they correspond to dwarf
galaxies with virial masses of $\approx 10^{10} \msun$ that fell into
the Milky Way $\gtrsim 9$ Gyr ago. Likewise, \citet{Be18} have
suggested that the highly radially anisotropic velocity distribution
of halo stars may be the imprint of a massive merger event, for which
evidence also exits in the radial profile of the stellar halo density
law ~\citep{De13}.

Retrograde substructures are interesting, because they may be related
to the anomalous globular cluster $\omega$~Centauri. This has a
present-day mass of $5 \times 10^6 M_\odot$~\citep{Me95} and is
believed to be the stripped nucleus of a dwarf
galaxy~\citep{Be03}. This is bolstered by the fact that
$\omega$~Centauri has long been known to contain multiple stellar
populations~\citep{No96,Su96,Be04}. Not merely do the stars in
$\omega$~Centauri exhibit a large metallicity spread~\citep{No95}, but
there are extreme star-to-star variations in many light
elements~\citep{Mar12,Mi17}. If $\omega$~Centauri was once a dwarf
galaxy, then its virial mass may have been as high as $10^{10}
M_\odot$ based on models of the chemical evolution of multi-population
clusters~\citep{Va11}. Dynamical evolutionary models find similar,
though somewhat lower, starting values of $\sim 10^8 - 10^9 M_\odot$
\citep[e.g.,][]{Be03,Ts03, Ts04}. Therefore, $\omega$~Centauri must
have disgorged much of its initial mass of stars (and dark matter) as
tidal debris in its passage to the inner Galaxy.

Searches for tidal debris in the solar neighbourhood date back to at
least \citet{Di02}, who found a retrograde signature in the solar
neighbourhood for stars in the metallicity range $-2.0 \le $[Fe/H]
$\le -1.5$. Further kinematic searches followed, though primarily with
small samples of stars concentrated in the solar
neighbourhood~\citep[e.g.,][]{Mi03, Br03, Me05, Fe15}.
\citet{Ma12} examined the line of sight velocities of $\sim 3000$
metal-poor stars within 5 kpc and conjectured that most of the
retrograde stars in the inner halo may be related to the disruption of
$\omega$~Centauri.  There have also been suggestions of evidence of
material torn from $\omega$~Centauri by \citet{Mo09} and \citet{He17}
based on their studies with 246 metal-poor stars and 1912 halo stars
respectively.  However, some specific groups that have been suggested
as likely contenders for material stripped off -- such as Kapteyn's
Moving Group~\citep{Wy10} and the so-called $\omega$~Centauri Moving
Group~\citep{Me05} -- have not survived detailed scrutiny based on the
chemical evidence~\citep{Na15}.

This paper is organised as follows.  Section 2 describes our algorithm
for substructure search in action space using the SDSS-\Gaia
catalogue. We identify 21 substructures in total with coherent
kinematics and narrow metallicity distributions.  Remarkably, we find
that some of the most significant substructures are comparatively
metal-rich, high energy and retrograde. Section 3 describes the
properties of our retrograde candidates, and uses simple models of
dynamical friction to investigate whether at least some of the new
retrograde substructures are likely to be the shards of
$\omega$~Centauri. We draw our conclusions in section 4.

 \begin{figure*}
 \begin{center}
\includegraphics[width=\textwidth]{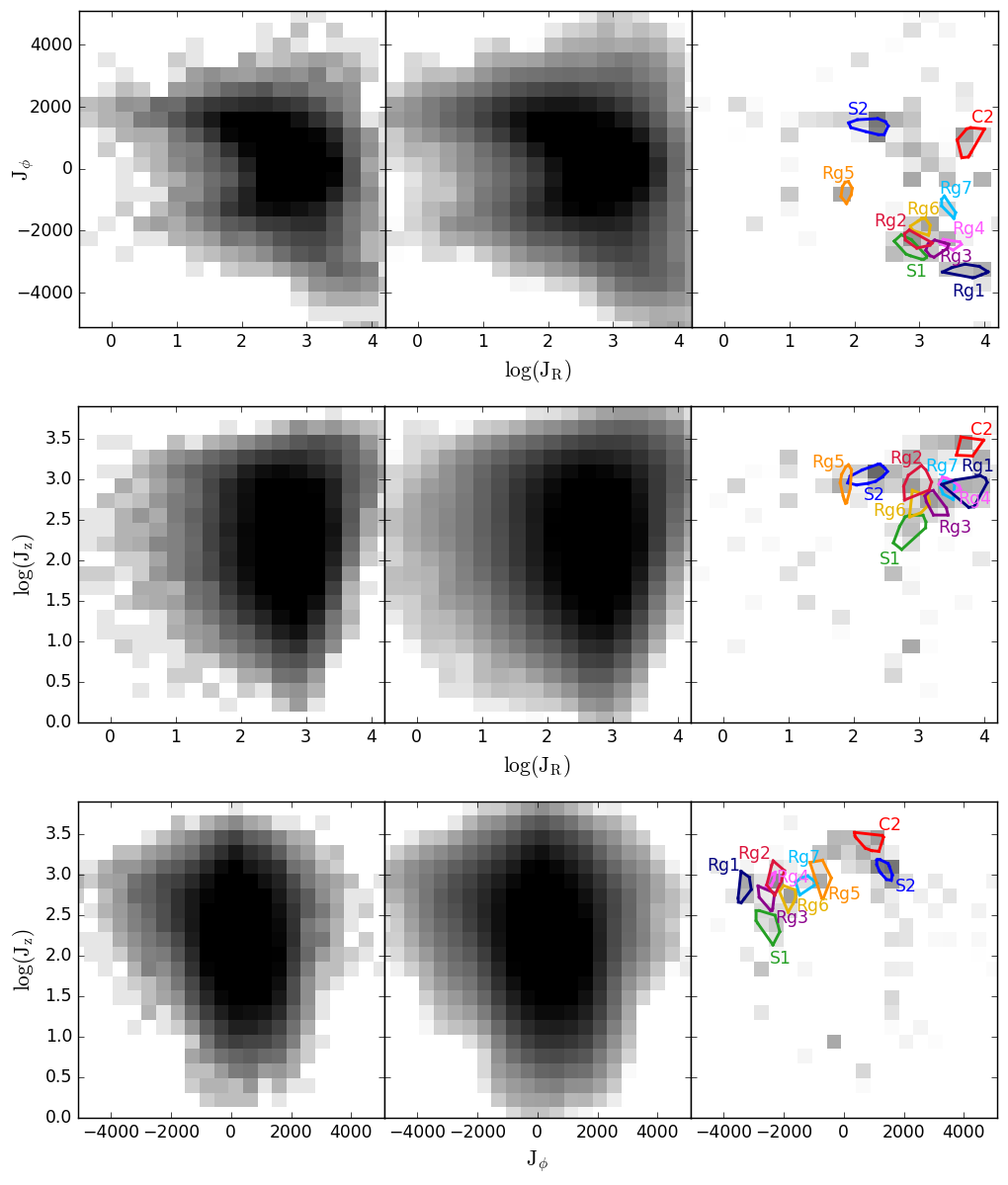}
 \end{center}
 \caption{Two-dimensional projection of the detection space.
   We show from left to right the data, the smooth Gaussian
   kernel density model, and the residuals. The rows show the
   principal planes in action space ($\log J_R,J_\phi$), ($\log J_R,
   \log J_z$) and ($J_\phi, \log J_z$) respectively. Reassuringly, the
   residuals correspond to the locations of the main pieces of
   substructure.}
 \label{fig:figonepfive}
\end{figure*}

 \begin{figure*}
   \begin{center}
 \includegraphics[width=\textwidth]{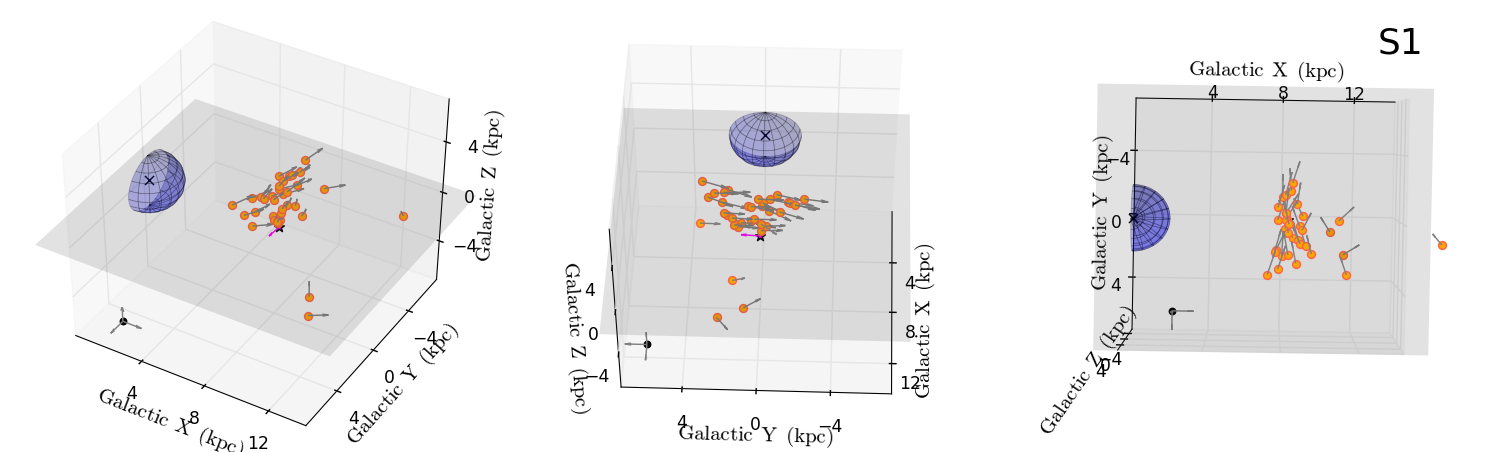}
 \includegraphics[width=\textwidth]{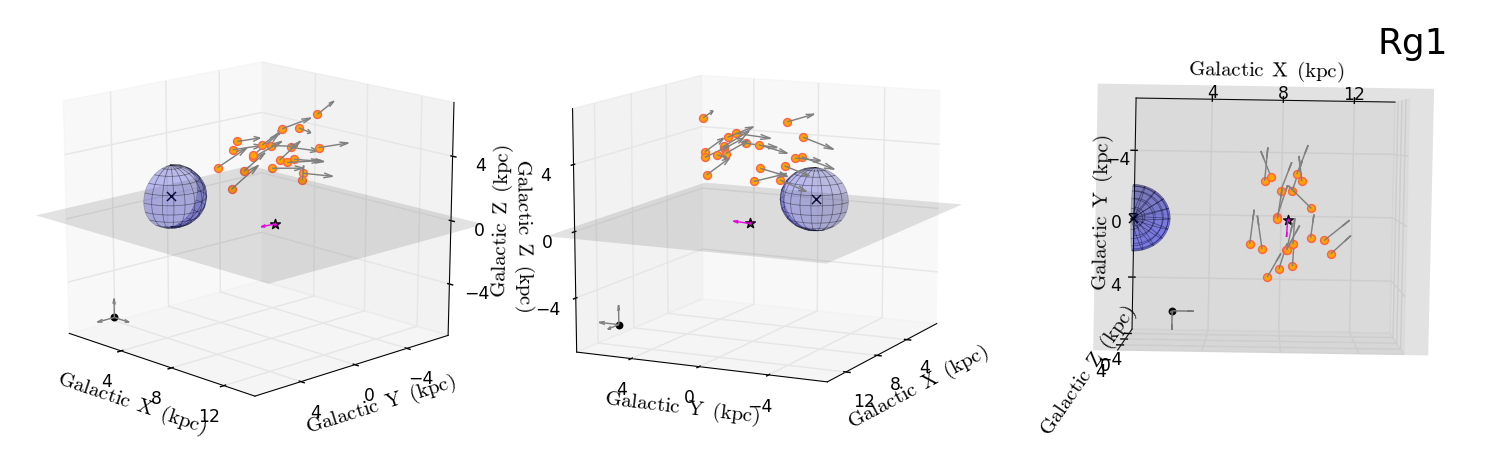}
 \includegraphics[width=\textwidth]{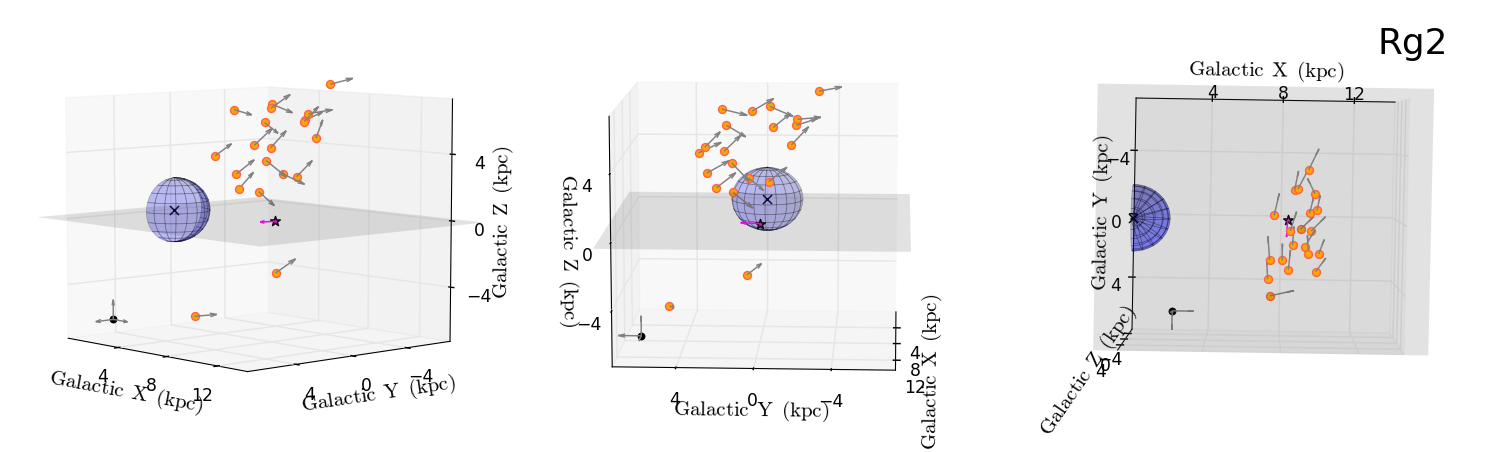}
 \includegraphics[width=\textwidth]{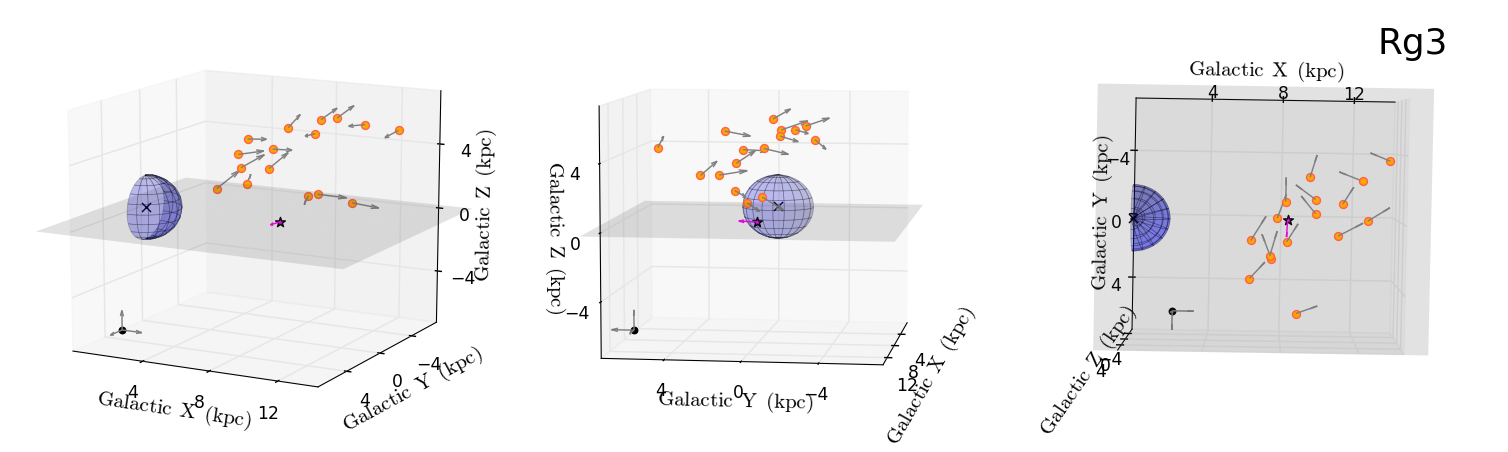}
 \end{center}
 \caption{Spatial distribution of 4 selected retrograde substructures,
   the previously known S1 and the new Rg1, Rg2, and Rg3. Left and
   Middle: Two views of the substructure depicting the overall shape
   and motion. Right: Projection of the substructure onto the Galactic
   plane. The arrow shows the total Galactocentric velocity. The Sun
   and the Sun's motion are marked as a star and a magenta arrow. A
   $2$\,kpc radius sphere and a grey plane are crude representation of
   the Galactic bulge and the Galactic plane to give a sense of
   substructure's scale and location in Galactic frame. A black triad
   of velocity vectors (scale of $300$\,km\,s$^{-1}$) is marked in
   each panel.}
 \label{fig:figtwo}
\end{figure*}

\begin{table*}

  \caption{The mean, mean absolute deviation and dispersions in
    positional and kinematic properties of the already known S1 and 7
    new retrograde candidate substructures (Rg1 -- Rg7). Also given
    are metallicity [Fe/H], as well as orbital properties, including
    energy $E$, circularity $\eta$ (or ratio of total angular momentum
    to the angular momentum of a circular orbit of the same energy),
    and orbital inclination $i$. An electronic list of substructure
    member stars is available from the authors.}

 \begin{center}
   \begin{tabular}{lrrrrrrrrc}
 \hline \hline
 \multicolumn{1}{c}{Name} &
 \multicolumn{1}{c}\null &
 \multicolumn{1}{c}{[Fe/H]} &
 \multicolumn{1}{c}{$(J_R,J_\phi,J_z)$} &
 \multicolumn{1}{c}{$\eta$} &
 \multicolumn{1}{c}{incl. $i$} &
 \multicolumn{1}{c}{$(X,Y,Z)$} &
 \multicolumn{1}{c}{$(v_R,v_\phi,v_z)$} &
 \multicolumn{1}{c}{E} &
 \multicolumn{1}{c}{Mem. no.} \\

 \multicolumn{1}{c}\null &
 \multicolumn{1}{c}\null &
 \multicolumn{1}{c}{\null} &
 \multicolumn{1}{c}{(kms$^{-1}kpc$)} &
 \multicolumn{1}{c}{\null} &
 \multicolumn{1}{c}{(deg)} &
 \multicolumn{1}{c}{(kpc)} &
 \multicolumn{1}{c}{(kms$^{-1}$)} &
 \multicolumn{1}{c}{(km$^2$s$^{-2}$)} &
 \multicolumn{1}{c}{\null}  \\

\hline

\null & Mean & -1.91 & (749.7,-2551.1,253.4) & 0.74 & 157.43 & (8.9,0.6,2.5) & (-8.6,-286.7,-67.9) & -118958 & \null \\
S1 & MAD & 0.22 & (186.6,180.9,51.2) & 0.04 & 2.25 & (1.0,1.2,1.2) & (104.5,40.9,45.8) & 5761 & 34 \\
\null & Dispersion & 0.26 & (234.8,210.4,60.3) & 0.05 & 2.56 & (1.6,1.4,1.9) & (115.3,49.9,60.0) & 6933 & \null \\
\hline\hline
\null & Mean & -1.60 & (6066.6,-3309.0,759.5) & 0.36 & 145.37 & (8.3,0.2,4.3) & (54.1,-393.8,68.3) & -65634 & \null \\
Rg1 & MAD & 0.14 & (2033.8,94.2,174.8) & 0.08 & 3.94 & (0.8,1.7,0.8) & (121.8,37.2,135.7) & 9459 & 20 \\
\null & Dispersion & 0.17 & (2452.3,116.3,197.0) & 0.09 & 4.54 & (1.1,1.9,1.0) & (147.2,45.7,149.4) & 11260 & \null \\
\hline
\null & Mean & -1.60 & (980.5,-2307.8,930.0) & 0.71 & 136.29 & (8.8,0.8,4.0) & (47.8,-254.6,71.8) & -109787 & \null \\
Rg2 & MAD & 0.12 & (270.5,109.3,175.2) & 0.05 & 3.56 & (0.7,1.7,2.4) & (91.5,22.6,149.4) & 6027 & 20 \\
\null & Dispersion & 0.11 & (303.1,140.8,221.0) & 0.06 & 4.47 & (0.8,2.0,3.2) & (110.2,27.8,172.1) & 6578 & \null \\
\hline
\null & Mean & -1.46 & (1844.3,-2550.5,503.5) & 0.54 & 148.35 & (9.3,0.1,4.2) & (28.0,-275.1,18.0) & -99899 & \null \\
Rg3 & MAD & 0.17 & (356.7,136.6,90.4) & 0.05 & 3.01 & (1.8,1.9,1.4) & (160.4,54.6,123.2) & 3978 & 16 \\
\null & Dispersion & 0.22 & (445.8,163.5,112.2) & 0.07 & 3.52 & (2.1,2.3,1.6) & (186.0,62.7,136.5) & 4769 & \null \\
\hline
\null & Mean & -1.47 & (3228.6,-2423.1,850.8) & 0.42 & 138.64 & (8.2,1.7,5.0) & (177.6,-287.9,117.9) & -84803 & \null \\
Rg4 & MAD & 0.14 & (683.5,64.8,89.8) & 0.06 & 2.40 & (1.1,1.5,1.4) & (88.5,35.5,142.1) & 5541 & 13 \\
\null & Dispersion & 0.19 & (781.9,85.2,111.4) & 0.07 & 3.03 & (1.6,1.8,1.7) & (114.6,46.0,162.7) & 6217 & \null \\
\hline
\null & Mean & -2.16 & (75.4,-723.5,937.4) & 0.89 & 114.98 & (8.2,0.3,1.7) & (-10.0,-82.3,-0.6) & -155848 & \null \\
Rg5 & MAD & 0.23 & (5.7,124.9,188.6) & 0.02 & 3.71 & (0.8,1.1,2.1) & (72.3,17.2,158.2) & 7073 & 29 \\
\null & Dispersion & 0.20 & (7.2,154.4,247.1) & 0.03 & 4.68 & (1.1,1.4,3.0) & (83.2,21.2,162.7) & 8588 & \null \\
\hline
\null & Mean & -1.63 & (1074.5,-1837.7,522.9) & 0.60 & 141.76 & (8.2,0.7,3.0) & (-11.9,-222.1,88.0) & -117935 & \null \\
Rg6 & MAD & 0.17 & (174.0,94.8,74.4) & 0.04 & 2.96 & (1.4,1.3,2.0) & (178.6,35.4,115.1) & 3336 & 30 \\
\null & Dispersion & 0.22 & (210.2,125.2,91.4) & 0.05 & 3.72 & (2.0,1.7,2.5) & (187.6,44.0,134.9) & 4439 & \null \\
\hline
\null & Mean & -1.48 & (2883.7,-1314.8,770.6) & 0.33 & 129.61 & (8.6,0.1,4.4) & (-92.3,-160.2,-33.4) & -95342 & \null \\
Rg7 & MAD & 0.17 & (385.4,148.8,70.7) & 0.02 & 3.24 & (1.6,1.1,2.0) & (255.2,53.2,138.0) & 5236 & 14 \\
\null & Dispersion & 0.24 & (447.6,190.7,96.6) & 0.03 & 4.07 & (2.0,1.4,2.6) & (274.6,66.5,178.1) & 6073 & \null \\
\hline

   \end{tabular}
 \end{center}
  \label{tab:basics_1}
\end{table*}

\begin{table*}

  \caption{The mean, mean absolute deviation and dispersions in
    positional and kinematic properties of the prograde or radial
    candidate substructures. This includes the previously known S2 and
    C2, as well as the new candidates (Cand8 -- Cand18). Also given
    are metallicity [Fe/H], as well as orbital properties, including
    energy $E$, circularity $\eta$ (or ratio of total angular momentum
    to the angular momentum of a circular orbit of the same energy),
    and orbital inclination $i$. A list of substructure member stars
    is available from the authors electronically.}

 \begin{center}
   \begin{tabular}{lrrrrrrrrc}
 \hline \hline
 \multicolumn{1}{c}{Name} &
 \multicolumn{1}{c}\null &
 \multicolumn{1}{c}{[Fe/H]} &
 \multicolumn{1}{c}{$(J_R,J_\phi,J_z)$} &
 \multicolumn{1}{c}{$\eta$} &
 \multicolumn{1}{c}{incl. $i$} &
 \multicolumn{1}{c}{$(X,Y,Z)$} &
 \multicolumn{1}{c}{$(v_R,v_\phi,v_z)$} &
 \multicolumn{1}{c}{E} &
 \multicolumn{1}{c}{Mem. no.} \\

 \multicolumn{1}{c}\null &
 \multicolumn{1}{c}\null &
 \multicolumn{1}{c}{\null} &
 \multicolumn{1}{c}{(kms$^{-1}kpc$)} &
 \multicolumn{1}{c}{\null} &
 \multicolumn{1}{c}{(deg)} &
 \multicolumn{1}{c}{(kpc)} &
 \multicolumn{1}{c}{(kms$^{-1}$)} &
 \multicolumn{1}{c}{(km$^2$s$^{-2}$)} &
 \multicolumn{1}{c}{\null}  \\

\hline
\null & Mean & -1.94 & (206.1,1363.3,1144.6) & 0.88 & 58.64 & (9.1,0.3,0.9) & (-11.2,159.6,-166.5) & -133500 & \null \\
S2 & MAD & 0.19 & (45.6,101.2,118.8) & 0.02 & 2.33 & (1.0,0.9,2.6) & (46.7,16.6,110.7) & 2488 & 73 \\
\null & Dispersion & 0.23 & (55.8,120.0,146.5) & 0.03 & 2.96 & (1.5,1.2,3.2) & (65.6,21.9,151.7) & 2987 & \null \\
\hline
\null & Mean & -1.45 & (5718.7,896.5,2577.4) & 0.30 & 75.55 & (9.0,-0.6,2.3) & (-242.4,109.9,180.7) & -67927 & \null \\
C2 & MAD & 0.11 & (938.3,247.9,282.7) & 0.03 & 3.66 & (0.9,0.8,1.9) & (155.1,28.5,189.5) & 4916 & 22 \\
\null & Dispersion & 0.13 & (1359.0,296.6,354.3) & 0.04 & 4.45 & (1.2,0.9,2.5) & (229.2,35.9,236.1) & 7011 & \null \\ \hline\hline
\null & Mean & -1.76 & (498.5,1695.6,940.0) & 0.77 & 51.72 & (9.3,0.2,1.9) & (8.2,192.0,-25.0) & -125103 & \null \\
Cand8 & MAD & 0.21 & (92.3,162.7,84.4) & 0.03 & 2.01 & (1.6,1.9,3.6) & (104.4,32.9,169.6) & 3537 & 49 \\
\null & Dispersion & 0.22 & (117.4,197.9,109.0) & 0.04 & 2.60 & (2.0,2.4,4.2) & (124.0,40.2,186.6) & 4317 & \null \\
\hline
\null & Mean & -1.82 & (672.2,1423.6,1488.1) & 0.75 & 61.75 & (9.1,-0.1,1.7) & (48.4,164.5,-171.1) & -118088 & \null \\
Cand9 & MAD & 0.15 & (182.1,101.2,150.0) & 0.04 & 1.90 & (0.9,1.2,2.5) & (118.3,18.0,125.1) & 4324 & 44 \\
\null & Dispersion & 0.18 & (221.4,123.2,201.3) & 0.05 & 2.40 & (1.2,1.6,3.4) & (139.6,24.2,172.0) & 5626 & \null \\
\hline
\null & Mean & -2.01 & (1127.1,94.2,2345.9) & 0.61 & 88.01 & (8.7,0.0,1.8) & (-115.6,17.4,169.3) & -115259 & \null \\
Cand10 & MAD & 0.13 & (255.7,123.3,269.8) & 0.04 & 2.73 & (1.4,1.0,2.0) & (175.1,15.1,137.8) & 6958 & 39 \\
\null & Dispersion & 0.18 & (306.7,144.5,329.2) & 0.06 & 3.23 & (1.9,1.3,2.8) & (200.3,18.5,197.5) & 7929 & \null \\
\hline
\null & Mean & -2.03 & (795.4,722.1,1903.9) & 0.69 & 74.93 & (9.1,-0.2,2.5) & (-82.8,91.7,41.4) & -119169 & \null \\
Cand11 & MAD & 0.20 & (113.3,181.1,139.5) & 0.04 & 2.68 & (1.7,0.9,2.5) & (153.8,26.0,221.2) & 4128 & 37 \\
\null & Dispersion & 0.13 & (144.3,222.8,185.1) & 0.04 & 3.33 & (2.1,1.2,3.2) & (171.1,36.6,239.8) & 4980 & \null \\
\hline
\null & Mean & -1.57 & (800.8,828.0,1505.6) & 0.66 & 70.26 & (9.6,-0.8,3.7) & (-33.3,94.4,60.9) & -122892 & \null \\
Cand12 & MAD & 0.17 & (174.1,152.3,129.1) & 0.05 & 2.46 & (1.6,2.1,3.2) & (151.8,21.9,173.4) & 4141 & 36 \\
\null & Dispersion & 0.19 & (204.5,197.8,153.0) & 0.06 & 3.01 & (2.0,2.5,4.2) & (175.4,26.8,197.5) & 5021 & \null \\
\hline
\null & Mean & -1.37 & (2272.7,-125.6,2356.9) & 0.44 & 92.40 & (9.1,1.0,3.1) & (-101.3,-9.3,98.4) & -98737 & \null \\
Cand13 & MAD & 0.13 & (360.1,183.8,373.0) & 0.04 & 3.71 & (1.9,1.6,3.2) & (223.1,22.0,203.0) & 5956 & 36 \\
\null & Dispersion & 0.18 & (481.6,239.5,449.8) & 0.06 & 4.88 & (2.5,2.0,4.1) & (246.4,30.4,238.1) & 7373 & \null \\
\hline
\null & Mean & -1.45 & (777.4,1837.4,219.4) & 0.65 & 30.97 & (11.5,0.2,1.6) & (-51.6,170.5,-7.5) & -128409 & \null \\
Cand14 & MAD & 0.15 & (91.2,130.2,37.5) & 0.04 & 1.36 & (1.6,1.4,3.1) & (155.1,20.9,79.9) & 2408 & 36 \\
\null & Dispersion & 0.17 & (116.0,169.3,44.3) & 0.05 & 1.75 & (1.9,1.8,3.6) & (166.6,27.0,90.1) & 3050 & \null \\
\hline
\null & Mean & -1.49 & (3041.5,1850.1,625.0) & 0.37 & 43.60 & (10.4,0.8,3.4) & (-73.6,186.4,-29.2) & -91662 & \null \\
Cand15 & MAD & 0.09 & (635.1,170.2,93.3) & 0.05 & 2.17 & (1.6,1.7,3.9) & (272.3,25.4,108.5) & 6905 & 19 \\
\null & Dispersion & 0.10 & (760.4,195.2,108.6) & 0.05 & 2.63 & (1.9,2.3,4.8) & (285.9,34.7,129.2) & 8796 & \null \\
\hline
\null & Mean & -1.43 & (2769.2,875.5,450.8) & 0.25 & 50.53 & (8.7,1.1,2.3) & (114.6,107.3,90.7) & -102594 & \null \\
Cand16 & MAD & 0.12 & (450.5,66.2,44.9) & 0.04 & 2.88 & (1.2,1.6,3.8) & (255.8,15.6,122.5) & 5501 & 17 \\
\null & Dispersion & 0.09 & (538.2,88.9,53.4) & 0.05 & 3.57 & (1.5,2.0,4.4) & (283.2,19.3,146.9) & 6526 & \null \\
\hline
\null & Mean & -2.13 & (1614.4,673.3,2263.4) & 0.56 & 77.22 & (9.7,-0.3,2.6) & (29.0,77.5,116.5) & -103328 & \null \\
Cand17 & MAD & 0.15 & (210.4,89.7,251.5) & 0.03 & 1.40 & (1.0,0.8,1.7) & (171.1,10.6,230.6) & 4013 & 14 \\
\null & Dispersion & 0.18 & (240.2,101.4,317.8) & 0.04 & 1.67 & (1.3,0.9,2.5) & (201.1,13.2,269.9) & 4778 & \null \\
\hline
\null & Mean & -1.27 & (8654.9,-18.1,1665.1) & 0.14 & 89.88 & (7.2,-0.1,5.3) & (-151.7,-8.3,-26.3) & -58877 & \null \\
Cand18 & MAD & 0.12 & (1060.5,223.2,294.1) & 0.02 & 6.96 & (2.3,1.6,1.4) & (398.6,35.0,117.5) & 4395 & 12 \\
\null & Dispersion & 0.14 & (1304.9,257.8,350.7) & 0.03 & 8.00 & (3.1,1.8,1.6) & (424.7,40.6,142.8) & 5216 & \null \\
\hline

   \end{tabular}
 \end{center}
  \label{tab:basics_2}
\end{table*}

\section{Detection of Substructures}

\subsection{Method}

We use the SDSS-{\it Gaia} catalogue. This is created by the
crossmatch between Gaia data release 1 (DR1), the Sloan Digital Sky
Survey data release 9 and LAMOST data release
3~\citep[see,][]{Ahn12,Lu15}.  Briefly, the main sequence turn-off
stars (MSTOs) are extracted using the cuts: extinction $\epsilon_r <
0.5$, $g,r,i$ magnitudes satisfying $14 < g < 20$, $14 < r < 20$, $14
< i < 20$, $0.2 < (g-r)_0 < 0.8$ with surface gravity $3.5 < \log g <
5.0$ and effective temperature $4500 < T_{\rm eff} < 8000$. The
rationale for the cuts is described in detail in \citet{Wi17}.  The blue
horizontal branch stars (BHBs) are chosen from $-0.25 < (g-r)_0 <
0.0$, $0.9 < (u-g)_0 < 1.4$ with spectroscopic parameters satisfying
$3.0 < \log g < 3.5$ and $8300 < T_{\rm eff} < 9300$. Photometric
parallaxes based on the SDSS photometry are used for MSTOs and BHBs
using the formulae in \citet{Iv08} and in \citet{De11} to give full
six-dimensional phase space coordinates.  We apply a series of quality
cuts to both the photometric and spectroscopic data to remove stars
with poor measurements as well as stars with a heliocentric radial
velocity error $> 15$ kms$^{-1}$, distance error $> 2.5$ kpc, and a
heliocentric distance $> 10$ kpc.

We then convert the observables to velocities in the Galactic rest-frame.
We use the Milky Way potential of \citet{Mc17}, which gives the
circular speed at the Sun as 232.8 km\,s$^{-1}$. For the Solar
peculiar motion, we use the most recent value from \citet{Sc10},
namely $(U,V,W) = (11.1, 12.24, 7.25)$ km\,s$^{-1}$.
The separation between the disk and the halo stars is carried out based on
their azimuthal velocity and their metallicity \citep[e.g.,][]{My18}.
The equation for the excision of disk stars is
\begin{equation}
  [{\rm Fe/H}] \gtrsim  -0.002 ~v_\phi - 0.9
\end{equation}
where $v_\phi$ is the azimuthal velocity in direction of the Milky Way
rotation and [Fe/H] the metallicity.  This equation gives a good
description of the more elaborate statistical separation displayed in
Figure 1 of \citet{My18}.  After the cuts, we obtain a sample of
62\,133 halo stars comprising 61\,911 MSTO stars and 222 BHB stars
(59\,811 stars with SDSS DR9 and 2\,322 stars with LAMOST DR3
spectroscopy). The locations of these stars are shown projected onto
the principal planes of the Galaxy in Fig.~\ref{fig:figzero}. Notice
that the sample extends well beyond the solar neighbourhood out to
heliocentric distances of 10 kpc. There are clear spatial selection
effects, and the footprint of the Sloan Digital Sky Survey can be
readily discerned. Nonetheless, the sample is kinematically unbiased
and has already proved to be a treasure trove for substructure
searches in velocity space~\citep{My18}.

Next, the actions of each star are computed using the numerical method
of \citet{Bi12} and \citet{Sa16}.  We construct a background model
representing the underlying smooth distribution of the data in the
3-dimensional action space $(\log(J_R),J_{\phi},\log(J_z))$. We
perform our search in logarithmic scale for $J_R$ and $J_z$ to
compensate for the increase in spread of $J_R$ and $J_z$ which can
reach large values for halo stars~\citep[see e.g., Figure 7 of
][]{Sa15}. The density estimation with a Gaussian kernel (KDE) from
{\tt Scikit-learn} \citep{Pe11} is used with the optimal bandwidth
determined by cross-validation.  From this model, we generate 200
random samples with the same size as the data. For each sampling, we
use a $k$-nearest neighbour search with $k=5$ to measure the density
at the location of each star in the actual data. The mean of these 200
independent measurements is considered as the local density $S_0$
expected by the model (computationally faster than deriving the model
density by Gaussian KDE itself).  The similar $k$-nearest neighbours
search is applied on the original data to obtain the actual measured
density $S$. From the probability density function, $P(S) \approx
S^{-k-1} \exp(-k S_0/S)$, we compute the probability percentile of the
measured density and convert it to the number of sigma indicating the
significance.

Stars with significance $>4$ are used as ``seeds'' for searching for
overdensities in action space. The seeds are first classified into
several groups based on their relative location in the action space by
the hierarchical agglomerative clustering implementation in {\tt
  Scikit-learn}.  For each seed, we collect nearby stars within a
local volume of ellipsoid with semi-axes corresponding to one third of
the standard deviation of each action. We discard any seeds that have
less than 5 stars within this volume. The collected stars are
classified by the Nearest Neighbours Classification from {\tt
  Scikit-learn}. The classifier is trained on the pre-classified
seeds and performs a distance-weighted $(k=3)$ neighbours
classification on stars. This provides us with a list of substructure
candidates.

For each candidate, we measure the volume of ellipsoid in action space
occupied by its member. The expected density (predicted by the model)
at the centre of this volume is used to estimate the expected number
of stars for the candidate, and hence obtain the significance (using
the same method as described above). To obtain a high quality list,
we require that a candidate (i) has significance $> 4$, (ii) contains
more than 10 stars and (iii) has a metallicity distribution function
(MDF) strongly peaked in comparison with the halo MDF. The latter is
judged by first decomposing the halo MDF into two Gaussians
\citep[with dispersions 0.38 and 0.27 as the result of Gaussian
  mixture model. see e.g., Figure 1 of ][]{My18}.  We require that a
Gaussian fit to our candidate MDF should have a dispersion less than
0.27, ensuring that it is peakier than the halo MDF. This gives us 21
candidate substructures.

\subsection{Algorithm Validation}

Before proceeding, we report two cross-checks. Using the public
software package AGAMA~\citep{Va18}, we generated a smooth model of a
stellar halo~\citep{Wi15} in the potential of \citet{Mc17}. We created
a catalogue of 250\,000 stars within a heliocentric distance of 10 kpc
around the Sun with the disk and the bulge region eliminated using
$|z|>1.5$ kpc and $r > 3.0$ kpc.  The algorithm identified no
substructures as it found no ``seeds''.

Secondly, we tested on publicly available stellar haloes created by
cosmological zoom-in simulations. We used the Aquarius catalogue
provided by ~\citet{Lo15}. The catalogue lists the "TreeID" for each
star providing the information of the parent satellite that brought
the star into the main halo. We obtained the catalogue of 250\,000
stars with 49 TreeIDs in the local volume of 10 kpc around the Sun
with disk and bulge region excluded. However, some of the TreeIDs
contribute very few stars in the local volume. There are 34 TreeIDs
with $> 50$ stars. This seems a reasonable figure against which to
measure performance of our method.

Our algorithm identified 28 candidate substructures after applying the
significance $\sigma > 4$ and number of member stars $> 10$ cuts. The
smallest candidate has 40 stars.  Although all the substructures
identified by the algorithm are real, there is not a one-to-one
correspondence between TreeIDs and candidates.  Of these, 2 TreeIDs
are detected as multiple candidates (4 candidates and 3 candidates
respectively) and 2 candidates show significant internal blending of
multiple TreeIDs.  For the case of blended candidates, we note that
these multiple TreeIDs in the same candidate have virtually the same
actions. Interestingly, they also occupy the same region in the
configuration space with the indistinguishable streaming motion --
therefore the same actions. This may be a case of multiple satellites
accreted to the main halo along the same dark matter filament at a
similar redshift. In this case, multiple TreeID groups are accreted
with almost identical kinematics.

We conclude that the present algorithm works well, in the sense of
identifying overdensities with high significance and generating
candidate lists for these overdensities. In particular, the tunable
parameters in the algorithm (bandwidth, linking procedure, number of
nearest neighbours) have been set conservatively. Although some highly
disrupted structures are missed, most substructures get picked up,
unless they are too small in size.

Our method has some points of similarity with \citet{He17} and also
some points of difference, which it is useful to summarise. Both
algorithms use the data to derive a smooth background model.  However,
our search proceeds in action space, whereas \citet{He17} use an \lq
integral of motion space' that is most appropriate for a spherical
potential. Secondly, \citet{He17} begins with a two-dimensional search
in ($E, J_\phi$) with a corroborating check for projections in the
third integral of motion, whereas we carry out our search in the
three-dimensional action space ($J_R, J_\phi, J_z$) from the
beginning.  Thirdly, \citet{He17} do not account for the Solar
peculiar motion and take the Local Standard of Rest as 220 kms$^{-1}$,
whereas we use the circular speed at the Sun as 232.8 km\,s$^{-1}$ and
the Solar peculiar motion from \citet{Sc10}, namely $(U,V,W) =
  (11.1, 12.24, 7.25)$ km\,s$^{-1}$. These differences are important,
as for very local stars they can cause a change from prograde to
retrograde.  Fourthly, we require that the metallicity distribution
functions of our substructures to be strongly peaked, whereas no such
requirement is imposed in \citet{He17}.  These differences in
methodology mean that we do not detect any of the ``VelHels''
identified by \citet{He17}. Many of the ``VelHels'' have rather broad
metallicity distribution functions~\citep{Ve18} and would fail our
criteria.

Although the algorithms are related, the main difference is the size
of the dataset through which we search. \citet{He17} uses a sample of
1912 halo stars extracted from TGAS crossmatched with RAVE. Our
algorithm has been applied to a sample of 62\,133 stars with full
six-dimensional phase space information in the SDSS-{\it Gaia}
catalogue~\citep[see e.g.,][]{De17,DeBoer18}. We identified 21 high
significance substructures. These all have morphological features that
resemble segments of orbits close to pericentre, as well as compact
metallicity distributions. The stars belonging to the substructures
are therefore kinematically and chemically similar.

 \begin{figure*}
 \begin{center}
 \includegraphics[width=\textwidth]{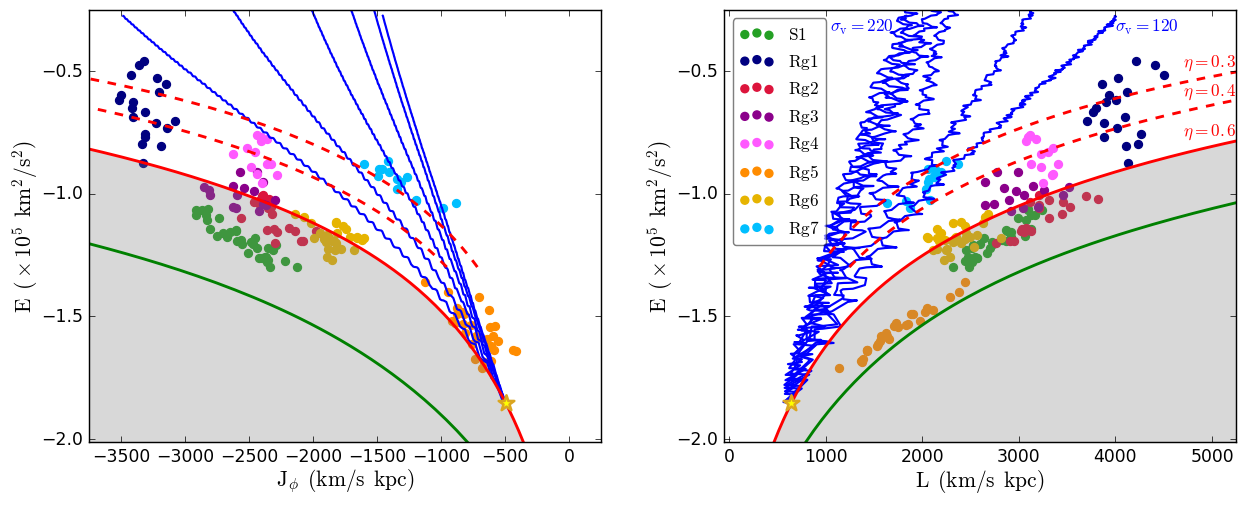}
 \end{center}
 \caption{Orbital tracks of $\omega$~Centauri in action space ($E,
   J_\phi$) and ($E,L$) as the progenitor sinks to its present
   location, together with the retrograde substructures (S1 and
   Rg1--7).  The golden star marks the present position of
   $\omega$~Centauri. The blue tracks the trajectory of the progenitor
   to the present-day $\omega$~Centauri, as given by numerical
   integration assuming Chandrasekhar dynamical friction with the
   velocity dispersion of the dark matter varying from 120 kms$^{-1}$
   to 220 kms$^{-1}$ in steps of 20 kms$^{-1}$.  We also show the
   evolution tracks of an object with a constant circularity $\eta
   \approx 0.6$ (solid red line) corresponding to $\omega$~Centauri
   today, whilst dashed red lines show further constant circularity
   tracks (0.4 and 0.3).  The grey shaded area shows the range of
   locations in which tidally-torn streams may not reside, as
   $\omega$~Centauri's circularity cannot have diminished during its
   orbital evolution. The green lines mark the (retrograde) circular
   orbit limit.  }
 \label{fig:figthree}
\end{figure*}

\section{Substructure Forensics}

Fig.~\ref{fig:figone} shows the 21 high significance substructures in
action space and integral of motion space. The data, the underlying
smooth model from the Gaussian kernel density estimator and the
residuals are shown in the left, middle and right panels of
Fig.~\ref{fig:figonepfive}. Reassuringly, the identified substructures
correspond to prominent residuals, mainly in the outer, relatively
less dense parts of the distribution. This is an effect due to the
imposition of high significance in candidate selection. Candidates
detected at the central denser regions are more
vulnerable to blending with random contaminants.  Since we use
compactness of the MDF of the candidates as one of the criteria for
validation, it is natural for us to identify more substructures with
high significance in less dense regions.  Another thing to notice is
that a significant number are in a retrograde tail of stars that
emanates from the main body of the distribution in
Fig.~\ref{fig:figone}. In fact, two of the top three most significant
substructures are retrograde. The population of high energy retrograde
stars provides a very happy hunting ground for halo substructure in
general.

We list the properties of all the retrograde candidates in
Table~\ref{tab:basics_1}.  The table gives their mean locations,
velocities in the Galactic rest-frame and metallicity. We also report
their orbital properties, including mean energy $E$ and circularity
$\eta$, which is the ratio of total angular momentum to the angular
momentum of a circular orbit of the same energy $L/L_{\rm
  circ}(E)$. Another orbital quantity of interest is the inclination
to the Galactic plane, defined as $i = \arccos (J_\phi/J)$ where $J$
is the absolute value of the total angular momentum.  Although we do
not study the new prograde substructure candidates in detail in this
paper, we list their basic properties in Table~\ref{tab:basics_2}.
Electronic tables of member stars are available from the authors.

\subsection{Cross-checks: Known Candidates (S1, S2, C2)}

\citet{My18} already identified six halo substructures in the
SDSS-{\it Gaia} catalogue from a search in velocity space. Only three
are recovered here with high significance, namely S1, S2 and C2. What
happened to the remaining substructures?  Two (S3, S4) are found, but
at lower significance than we imposed here. C1 is also identified with
a large portion of new members, but it fails the requirement that we
insisted that the substructure have a compact metallicity
distribution. The velocity-based search was more successful in
recovering seemingly clean stream-like structures for S2 and
C2. However, the power of action space is that it can associate
patches of substructures from different pericentric
passages~\citep{He99}. This means that more highly phase-mixed
material can be associated with the substructure, and so we expect
more disrupted morphologies.

\subsection{The Retrograde Candidates}

These include the previously known S1, and the seven new retrograde
candidates (Rg1--7, in order of decreasing significance). The
morphology of some of the retrograde substructures is shown in
Fig.~\ref{fig:figtwo}.  Their shapes are strongly affected by the
footprint, as the stars must lie in the SDSS so the coverage of the
Southern Galactic hemisphere is patchy. Occasionally, there are stars
that do not seem to agree with the overall morphology of the
substructure (e.g., in Rg3 there are two stars whose velocity vectors
run counter to the trends seen in the remaining stars in the
arm). These could be contaminants, but they could also be phase-mixed
material.  Nonetheless, the overall shapes of the substructures, as
well as their velocity distributions, are consistent with orbital
segments close to pericentre. The candidates all share similar
characteristics in that they are retrograde and all (but one) belong
to \citet{MyPre}'s categorisation of the comparatively metal-rich halo
($-1.9 <$ [Fe/H] $< -1.3$). They are tightly clustered in azimuthal
action $J_\phi$, but typically have much larger spreads in $J_R$ and
$J_z$.  It is interesting to compare S1 as selected in action space
with the more ragged view of the same substructure as selected in
velocity space and given in Figure 5 of \citet{My18}. This retrograde
substructure passes right through the solar neighbourhood. If there is
a dark matter stream associated with this substructure, then it it may
have important consequences for direct detection experiments.

Of course, $\omega$~Centauri is known to be on a retrograde orbit.
Its proper motion has recently been re-measured by \citet{Li18} and
differs somewhat from the previous value.  Using the potential of
\citet{Mc17}, the present energy of $\omega$~Centauri is $-1.85 \times
10^5$ km$^2$ s$^2$, whilst its actions ($J_R, J_\phi, J_z$) are
(264.5, -496.4, 93.5) kms$^{-1}$ kpc.  Its position is marked as a
golden star in the action plots of Fig.~\ref{fig:figone}.  This gives
a total angular momentum of 646.62 kms$^{-1}$ kpc and a present day
circularity of $\eta \approx 0.60$ for $\omega$~Centauri. Usually, the
effect of dynamical friction on orbits of moderate eccentricity is to
circularize orbits. However, \citet{Bo99} find that the orbital
circularity can sometimes stay roughly constant throughout the decay.
The eccentricity decreases near the pericentre, but increases near the
apocentre such that there is only mild net circularisation or
radialisation in their simulations in an admittedly spherical
potential \citep[see Figure 9 of ][]{Bo99}. It is reasonable to
conjecture that the orbit of $\omega$~Centauri can only get more
circular with time, or -- in this limiting case -- stay
constant. Thus, the circularity $\eta =0.6$ line is a limit below
which it is not sensible to associate substructure with
$\omega$~Centauri.  This rules out S1, Rg2 and Rg5 as belonging to the
sinking $\omega$ Centauri.

The circularity $\eta =0.6$ line is shown in Fig~\ref{fig:figthree}
with the region below it shaded grey as forbidden.  We also show the
tracks in red for objects evolving with constant circularity of $0.4$
and $0.5$ in action space.  In addition, we have supplemented these
with blue tracks showing the simple model trajectory of an
$\omega$~Centauri progenitor (represented as a point mass of $5 \times
10^8 \msun$) moving in the Galactic potential of \citet{Mc17} and
under the influence of dynamical friction as judged by the
\citet{Ch43} formula, with the velocity dispersion of the dark matter
particles as $120$ kms$^{-1}$ to 220 kms$^{-1}$ in 20 kms$^{-1}$
intervals~\citep[see also, Chapter 8.1 of][]{BT08}. We use the factor
$\Lambda$ in the Coulomb logarithm from the equation (8.1b) in
\citet{BT08}.  We note that these tracks are simple model trajectories
and although the Chandrasekhar formula can provide a good description
for orbital decay under dynamical friction \citep{BT08}, a more
realistic picture will require more sophisticated methods such as
N-body simulations~\citep[see also,][]{We89,Fu06}.  Although we
consider $\omega$~Centauri as a point mass, its internal velocity
dispersion could produce scatter about the tracks.  Still, at $5
\times 10^8 \msun$, the scatter would have a modest effect on the
overall direction of the trajectory. The rate of circularisation does
depend on the choice of parameters, especially the velocity dispersion
of the halo.  These tracks are much steeper, but it is actually
difficult to push the trajectories to lower values of $J_\phi$ than
that of the present day $\omega$~Centauri.  Of course, this
calculation omits any effects due to mass loss from $\omega$~Centauri
or evolution of the Milky Way potential.  Given that the structure of
the progenitor and the workings of dynamical friction in the Galaxy
are not well-known, we regard the region between the constant
circularity line $\eta = 0.60$ and the most extreme Chandrasekhar
curve as the likely area in which tidal fragments are to be sought and
found. This suggests that the substructures Rg1, Rg3, Rg4, Rg6 and Rg7
are all possible candidates.

Further evidence can be provided by the inclinations of the
substructures, which are listed in Table~\ref{tab:basics_1}.  Here, we
use the traditional convention that inclinations greater than
$90^\circ$ describe retrograde orbits.  The effect of dynamical
friction is to drag the orbit of a sinking satellite down towards the
Galactic plane. Unsurprisingly $\omega$~Centauri is now on a rather
low inclination orbit $i_{\rm OC} = 140.15^\circ$. So, candidates with
more inclined retrograde orbits (that is, smaller $i$), or within the
range of their dispersion with the present day inclination of
$\omega$~Centauri, are feasible.  Rg1 has a slightly less inclined
orbit ($i = 145.37^\circ$), but considering its dispersion of
$4.54^\circ$, it is still plausible.  Rg3 has a less inclined orbit
($i = 148.35^\circ$), and even taking into account its dispersion, it
does not cause it to overlap with $i_{\rm OC}$. Rg4 has comparable but
more inclined orbit ($138.64^\circ$).  Rg6 has a slightly less
inclined orbit ($141.76^\circ$), while its dispersion takes it within
the range.  Rg7 has considerably more inclined orbit ($129.61^\circ$)
-- more than $10^\circ$ difference. This leaves Rg1, Rg4, Rg6 as the
strongest candidates, with Rg3 and Rg7 somewhat less favoured.

The validity of the claims can be established by seeing which
substructures are chemically consistent with $\omega$~Centauri via
high resolution spectroscopy. \citet{Na15} studied two prominent
pieces of retrograde substructure, Kapteyn's Moving Group, and the
so-called $\omega$~Centauri group. Both have been previously been
claimed to be material shed by $\omega$~Centauri on its journey to the
inner Galaxy~\citep{Me05,Wy10}. However, both groups are not related
to $\omega$~Centauri, based on abundances from Na, O, Mg, Al, Ca and
Ba derived from optical spectra.  In particular, $\omega$~Centauri has
characteristic Na-O and Mg-Al patterns of abundances for moderately
metal-rich halo stars, as well as an overabundance of Ba, that are
different from the halo field stars. The GALAH survey~\citep{Bu18},
with its range of elemental abundances, may also be useful here.

If the substructures are not related to $\omega$~Centauri, then they
are perhaps even more interesting and puzzling!  Presumably they must
then be the remnants of objects that are highly phase-mixed and so
little now remains even of the nucleus.  Studying the elemental
abundance ratios of the retrograde substructure will greatly benefit
the unravelling of their true origin. In particular, we would obtain
evidence on the importance of rapid (r) and slow (s) process
enrichment. It would be interesting to see if they show evidence for
the anomalous r-process enhancement, already detected for some of the
faintest dwarf galaxies~\citep{Ji16, Ro17}. To this end, the authors
happily make available electronic tables of the member stars in the
retrograde substructures as target lists for spectroscopy.

\section{Conclusions}

This paper has developed a new algorithm to search for substructure in
action space.  As actions are conserved under slow evolution of the
potential, stars accreted onto the Milky Way halo in the same merger
event should be clustered in action space. Thus, the algorithm
searches for significant overdensities with respect to the
data-derived background model. The metallicity distribution function
of the substructures is required to be more strongly peaked than the
stellar halo metallicity distribution function itself. The final
substructure candidates are therefore clustered both in action and in
metallicity.  Our algorithm has been validated against mock catalogues
of substructure in the Aquarius cosmological zoom-in simulations
provided by ~\citet{Lo15}.

This algorithm is similar in spirit to our earlier search strategy in
velocity space, though here we have used a Kernel density estimator to
model the background rather than a Gaussian mixture
model~\citep{My18}. We applied our algorithm to a sample of 62\,133
halo stars with full phase space coordinates extracted from the
SDSS-\Gaia catalogue. The sample size is at least an order of
magnitude greater than any previous substructure search in phase
space~\citep[see e.g.,][]{Mo09,He17}. The stars extend out to
heliocentric distances of $\sim 10$ kpc, and this permits us to
identify coherent features in phase space in an unprecedently large
volume of the Galaxy.

Altogether, we identified 21 high significance substructures in action
space.  Here, we have focussed on eight substructures that lie in the
retrograde, high energy portion of action space. This includes the
previously discovered S1 substructure~\citep{My18}, as well as seven
new candidates (Rg1--7). \citet{MyPre} already showed that the
retrograde, high energy stars in the local halo are confined to a
restricted range of metallicities ($-1.9 <$ [Fe/H] $< -1.3$). The
origin of this high energy and clumpy component of the local stellar
halo remains a puzzle. Although the substructure must have come from
mergers of retrograde satellites, it remains unclear whether one large
satellite or multiple smaller ones are responsible.

One possible source of the abundant retrograde substructure is the
anomalous globular cluster, $\omega$~Centauri. There is a long history
of searches in the solar neighbourhood for stars tidally torn from
$\omega$~Centauri \citep[e.g.,][]{Di02,Me05,Mo09}. On studying a
sample of metal-poor halo giants within $\sim 5$ kpc, \citet{Ma12}
made the bold conjecture that the disruption of the progenitor of
$\omega$~Centauri may have generated a very substantial part of the
retrograde population in the stellar halo. It is this hypothesis that
we can hope to test with substructure searches in deeper halo
catalogues like SDSS-\Gaia.

Here, we have shown based on kinematic evidence that three of our
substructures (Rg1, Rg4, Rg6) could be the shards of
$\omega$~Centauri. Rg3 and Rg7 are also possible, though they are
somewhat disfavoured on the grounds of their present inclination. S1,
Rg2 and Rg5 seem ruled out on the grounds of their present
circularity. The timescale of the orbital decay due to the dynamical
friction depends on the mass of the satellite
\citep[e.g.][]{Bo99}. Since this timescale must be shorter than a
Hubble time, then, given the current energy and location of
$\omega$~Centauri, the progenitor must have had a mass of at least $5
\times 10^8 \msun$, comparable to the value found by \citet{Be03}.
This sets a lower bound, as this is an average mass throughout the
orbital decay over the Hubble time. Moreover, the mass loss from the
tidal stripping and the evolution of the Milky Way potential could
cause the actual initial mass to be greater by perhaps an order of
magnitude~\citep[e.g.,][]{Ts03}.

The most direct way to test the claims of this paper is by obtaining
high resolution spectroscopy of the candidate stars in the
substructures.  In particular, $\omega$~Centauri has characteristic
Na-O and Mg-Al patterns of abundances for moderately metal-rich halo
stars, as well as an overabundance of Ba, that are different from the
halo field stars~\citep[c.f.][]{Na15}. Furthermore, suppose for
example we establish that Rg3 and Rg4 (but not the others) were
associated with $\omega$~Centauri. Then, this would provide
significant constraints on the progenitor and the action of dynamical
friction, as we would know whether the orbit is circularising.
Another intriguing possibility is that the highest energy
substructures may have been stripped before extended star formation
and multiple population enrichment, and so it may even be possible to
see gradients across the substructures.

If chemical evidence disproves our assertion that some of the
retrograde substructures belong to $\omega$~Centauri, then the
situation is perhaps even more interesting. It leaves us with two
major puzzles. First, where are the substantial amounts of debris that
must have been shed by the $\omega$~Centauri progenitor? And second,
what is the origin of the high energy, retrograde halo which is riven
with substructure?  The recent release of the Gaia DR2
dataset~\citep{GaDR2_sum} offers further golden prospects for
resolving these puzzles, as well as for harnessing the power of
substructure identification algorithms to build a complete inventory
of merger remnants in the stellar halo.  The algorithms and techniques
that we have developed here will have no small part to play.

\section*{acknowledgements}
We thank the Cambridge Streams Group (especially Eugene Vasiliev) and
Nicola Amorisco for helpful conversations.  GCM thanks Boustany
Foundation, Cambridge Commonwealth, European \& International Trust
and Isaac Newton Studentship for their support on his work. JLS thanks
the Science and Technology Facilities Council for financial
support. We are grateful to the anonymous referee who helped us improve
this work. The research leading to these results has received partial
support from the European Research Council under the European Union's
Seventh Framework Programme (FP/2007-2013) / ERC Grant Agreement
no. 308024.

\label{lastpage}
\end{document}